\documentclass[usegraphicx,useAMS]{mn2e}
\title[A non-main-sequence secondary in SY Cancri]
{A non-main-sequence secondary in SY Cancri}
\author[R.C. Smith, O. Mehes, D. Vande Putte and N.A. Hawkins]
{Robert Connon Smith\thanks{E-mail: rcs@sussex.ac.uk}, Otto Mehes, 
Dave Vande Putte,
Nigel A. Hawkins
\\
Astronomy Centre, Department of Physics and Astronomy, University of Sussex,
Falmer, Brighton BN1 9QH, UK}
\date{Accepted 2005 March 22.\hspace{3mm}
      Received 2005 February 17;\hspace{3mm}
      in original form 2004 August 4}

\def\kms{\thinspace\hbox{$\hbox{km}\thinspace\hbox{s}^{-1}$}} 
\def\msun{\thinspace\hbox{M$_{\odot}$}} 
\def\deg{\hbox{$^\circ$}} 
\newcommand{\nai}{Na\,{\small I}} 
\newcommand{\oxi}{O{\small I}} 
\newcommand{\ea}{{et al.}\ }

\begin{document}

\maketitle

 \begin{abstract}
Simultaneous spectroscopic and photometric observations of the Z Cam type 
dwarf nova SY Cancri were used to obtain absolute flux calibrations. A 
comparison of the photometric calibration with a wide slit spectrophotometric 
calibration showed that either method is equally satisfactory. A radial 
velocity study of the secondary star, made using the far red \nai\ doublet, 
yielded a semi-amplitude of $K_2 = 127\pm23$\,km\,s$^{-1}$. Taking the 
published value of $86\pm9$\,km\,s$^{-1}$ for $K_1$ gives a mass ratio 
$q = M_2/M_1 = 0.68 \pm 0.14$; this is very different from the value of 
$1.13 \pm 0.35$ quoted in the literature. Using the new lower mass ratio, 
and constraining the mass of the white dwarf to be within reasonable limits, 
then leads to a mass for the secondary star that is substantially less than 
would be expected for its orbital period if it satisfied a main-sequence 
mass-radius relationship. We find a spectral type of M0 that is consistent 
with that expected for a main-sequence star of the low mass we have found. 
However, in order to fill its Roche lobe, the secondary must be
significantly larger than a main sequence star of that mass and spectral type. 
The secondary is definitely not a normal main-sequence star.
\end{abstract} 

\section{Introduction}

The cataclysmic variable (CV) SY~Cancri is listed in the Ritter~\&\ Kolb
\nocite{ritter03} (2003) catalogue of CVs as a Z~Cam type dwarf nova with a period
of $0.380\pm0.001$ days ($9.13\pm0.024$ hours). In common with most Z~Cam stars
(Warner~1995, p.165), \nocite{warner95} there are no eclipses, the published
inclination being $26\deg\pm6\deg$ (Shafter~1983). \nocite{shafter83} An
intriguing result is the mass ratio, which is given by Shafter (1983) as $ M_2/M_1
= 1.13\pm0.35$. Despite the large uncertainty, this suggests that the secondary
star could be the more massive of the pair. If true, this would be rather
surprising, since the canonical picture of mass transfer in CVs assumes that the
secondary star is less massive than the white dwarf. Mass transfer can then
proceed on a slow timescale determined by the rate of angular momentum loss from
the system (e.g. A. King 1988). \nocite{aking88} If the secondary is the more
massive component, it should be transferring mass on a {\em dynamical} timescale
(A. King 1988). There is no evidence for that in SY~Cnc. Since we made our
observations, more detailed calculations by Politano (1996)\nocite{politano96}
have shown that the situation is more complicated than this, and estimates from
his Figure 2 show that the system would be stable with this mass ratio if the mass
of the secondary in solar masses were in the range 0.75 to 1.30. None the less, 
the Shafter mass ratio is unusually large and, as we shall see, stability 
criteria do play a role in determining the parameters of the system.

\begin{table*}
\begin{center}
\begin{tabular}{|c|c|c|c|c|c|c|l|}
\hline
 Night & Date &  UT   & Exposure &      HJD     & Skyshift & Heliocentric & Notes \\
 No. & 1989 & Start &   Time   & mid-exposure &          & Correction   & \\
& April & Time  &  (secs)  & (-2447600.0) &  (km\,s$^{-1}$)  &   (km\,s$^{-1}$)     &
\\
\hline 1 & 13/14 & 20:50 & 1024 & 30.3762728 &  0.00000 & 0.03 & R831R grating. \\

    &  & 21:12 & 1024 & 30.3915678 &  0.00000 & 0.06 & \\

3 & 15/16 & 22:19 & 1024 & 32.4379346 &  0.00000 & 0.19 & \\

    &  & 22:58 & 1024 & 32.4651007 &  0.00000 & 0.24 & \\

    & & 23:17 & 1024 & 32.4779764 &  3.98127 & 0.26 & \\

     & & 23:37 & 1024 & 32.4913055 &  0.00000 & 0.28 & \\

     & & 23:58 &  512 & 32.5039073 &  0.00000 & 0.31 & Wide slit. \\

     & & 00:11 &  512 & 32.5128255 &  0.61694 & 0.32 & Wide slit. \\

4 & 16/17 & 21:27 & 1024 & 33.4014396 &  0.00000 & 0.11 & R400R grating from here
on.\\

5 & 17/18 & 20:27 & 1024 & 34.3591914 &  0.00000 & 0.01 & \\

    &  & 20:49 & 1024 & 34.3752031 &  0.00000 & 0.05 & \\

    &  & 21:09 & 1024 & 34.3881816 & -1.98700 & 0.08 & \\

    &  &       & 1024 & 34.4005943 & -3.11917 & 0.11 & \\

    &  &       & 1024 & 34.4130141 & -8.13713 & 0.14 & \\

    &  &       & 1024 & 34.4254154 & 34.13813 & 0.17 & \\

    &  &       & 1024 & 34.4378249 &  0.00000 & 0.20 & \\

    &  & 22:41 & 1024 & 34.4528655 &  0.00000 & 0.23 & \\

    &  &       & 1024 & 34.4652682 &  3.15828 & 0.25 & \\

    &  &       & 1024 & 34.4776773 &  0.76570 & 0.27 & \\

    &  & 23:40 & 1024 & 34.4938585 &  0.41650 & 0.30 & Wide slit. \\

    &  & 00:03 & 1024 & 34.5096814 &  0.00000 & 0.32 & Wide slit. \\

6 & 18/19 & 20:46 & 1024 & 35.3729502 &  0.00000 & 0.05 & \\

    &  &       & 1024 & 35.3853784 &  0.92120 & 0.08 & \\

    &  &       & 1024 & 35.3978143 &  6.25017 & 0.11 & \\

    &  &       & 1024 & 35.4102357 & -9.07470 & 0.14 & \\

    &  &       & 1024 & 35.4226507 & -3.83577 & 0.17 & \\

    &  &       & 1024 & 35.4350579 &  0.00000 & 0.19 & \\

    &  & 22:44 & 1024 & 35.4547941 &  0.00000 & 0.23 & \\

    &  &       & 1024 & 35.4672223 &  3.25285 & 0.25 & \\

    &  &       & 1024 & 35.4796619 &  9.16655 & 0.28 & \\

    &  &       & 1024 & 35.4921105 & -0.76024 & 0.30 & \\

    &  &       & 1024 & 35.5045500 &  0.00000 & 0.31 & \\

7 & 19/20 & 21:23 & 1024 & 36.3985788 &  0.00000 & 0.12 & \\

    &  &       & 1024 & 36.4110005 &  2.40349 & 0.15 & \\

    &  &       & 1024 & 36.4234270 &  5.71727 & 0.17 & \\

    &  &       & 1024 & 36.4358591 &  3.62479 & 0.20 & \\

    &  &       & 1024 & 36.4482748 &  6.37748 & 0.23 & \\

    &  & 22:55 & 1024 & 36.4624989 & -1.04306 & 0.26 & \\

    &  &       & 1024 & 36.4749092 & -1.53231 & 0.28 & \\

    &  &       & 1024 & 36.4873283 & -1.17674 & 0.30 & \\

    &  &       & 1024 & 36.4997447 &  0.00000 & 0.32 & \\

    &  & 00:09 & 1024 & 36.5140711 &  0.00000 & 0.33 & \\

\hline
\end{tabular}
\caption[]{Journal of spectroscopic observations of SY Cnc. Most of the
observations were taken at the lower resolution, to obtain wider wavelength
coverage. No observations of SY Cnc were taken on night 2 (April 14/15)}
\label{syj:spec}
\end{center}
\end{table*}

Shafter's mass ratio estimate is based on detection of emission lines only,
together with an empirical calibration (Shafter~1983) of a relationship between
the mass ratio and the width of the H$\alpha$ {\em emission} line, interpreted as
a rotation velocity in the disc. It seemed more plausible that this relationship
is violated in SY Cnc than that the secondary is really the more massive
component, but only a direct detection of the absorption line spectrum of the
secondary star could verify that. We therefore decided to try to make a direct
measurement of $K_{2}$ to check the mass ratio and test the assumption made by
Shafter. We planned to use the \nai\ doublet near 8200\,\AA, which had been
successfully used to measure radial velocities for other secondary components
(Friend \ea 1990a, b). \nocite{friend90a}\nocite{friend90b}

The long period suggests that the secondary should be relatively massive
($\sim$1\,M$_{\odot}$) if it is on the main sequence and fills its Roche lobe, 
as is commonly the case for CV secondaries (Smith \&\ Dhillon 1998). This mass
corresponds to an early to mid G type star (which is consistent with a recent
spectral type estimate of G2 from IR spectra -- Harrison et al. 2004). 
However, Patterson~(1981) \nocite{patterson81}
reported a late G type spectrum, while Szkody~(1981) \nocite{szkody81} said that a
secondary with an effective temperature of $\simeq4500$K was needed to fit the UV,
optical and IR spectra, which suggested an early K type spectrum. These
discrepancies already suggested that the secondary is not quite normal. The rather
early spectral type meant that we would not expect the  \nai\ doublet to be easily
detectable, but Friend found a possible detection in his work (see Friend {et
al}~1988). \nocite{friend88} The system was in outburst at that time, making the
detection harder, so we hoped to be able to obtain a good radial velocity curve if
we found the system in quiescence. In practice, the system appeared to be on
decline from outburst, but the \nai\ doublet was still both detectable and
measurable.

\section{Observations and data reduction}

\subsection{Observations}

\subsubsection{Spectroscopy}

\begin{table*}
\begin{center}
\begin{tabular}{|c|c|c|c|c|p{2.2cm}|}
\hline
 System & Date & Spectral Type & RV & Helio. Corr. & Notes \\
\hline
%
%
%

Gl 673  & 19/20 & K7V   &   95.66 &  0.10 & 2.304 \AA /pixel \\

Gl 488  & 18/19 & M0.5e &   27.69 &  0.08 & \\

Gl 205  & 17/18 & M1    &    8.52 &  0.31 & \\

Gl 393  & 16/17 & M2    &    8.36 &  0.29 & \\

       & 17/18 &       &    8.36 &  0.28 & \\

Gl 701  & 17/18 & M2    &   32.47 &  0.07 & \\

Gl 752A & 16/17 & M3    &   35.82 & -0.08 & \\

Gl 654  & 15/16 & M3.5  &   34.60 &  0.15 & \\

Gl 555  & 19/20 & M4    &   28.62 & -0.12 & \\

Gl 285  & 16/17 & M4.5  &   26.59 &  0.11 & \\

Gl 699  & 15/16 & M5    & -110.86 &  0.07 & \\

Gl 473  & 18/19 & M5.5e &   15.50 &  0.10 & \\

\hline
\end{tabular}
\caption[]{Red dwarf radial velocity standards observed at low resolution on the
different nights in 1989 April.} \label{syj:red}
\end{center}
\end{table*}

\begin{table*}
\begin{center}
\begin{tabular}{|c|c|r@{--}l|c|c|l|}
\hline

Night & Date & \multicolumn{2}{c|}{Time} & Number of &  Exposure   & Comments \\

No. & 1989 April &  \multicolumn{2}{c|}{UT}  & Exposures & Time (secs) & \\

\hline

1 & 13/14 & 21:02 & 21:40 &   6 & various & UBVRIZ colours only. \\

4 & 16/17 & 21:13 & 21:16 &   1 & 128    & I band. \\

5 & 17/18 & 20:52 & 20:59 &   2 &  64    & I \\

  &    & 20:59 & 21:29 &  30 &  32    & I \\

  &    & 21:29 & 21:37 &   4 & various & UBVR colours. \\

  &    & 21:52 & 00:34 & 105 &  32    & I band. Brief gap for standard. \\

6 & 18/19 & 21:21 & 00:15 & 167 &  32    & I \\

  &     & 00:28 & 00:33 &   4 & various & UBVR colours. \\

7 & 19/20 & 20:39 & 21:45 &  62 &  32    & I \\

  &     & 21:45 & 21:52 &   4 & various & UBVR colours. \\

  &     & 22:12 & 22:53 &  32 &  32    & I \\

  &     & 23:15 & 00:45 &  71 &  32    & I band. Brief gap for standard. \\

\hline
\end{tabular}
\caption[]{Journal of all photometric observations of SY Cancri. The table does
not include several points which were unusable for various reasons. The most
common reasons were the frame being saturated, the telescope moving during an
exposure, or software problems causing frames to be either lost or corrupted. The
single I-band exposures taken on nights 1 and 4 were not used.} \label{syj:phot}
\end{center}
\end{table*}

We obtained 42 spectra of SY Cnc 
spread over six nights in 1989 April, as part of a more extensive observing
programme using the Intermediate Dispersion Spectrograph on the Isaac Newton
Telescope at the Roque de los Muchachos Observatory on the island of La Palma. We
used the 235\,mm camera and a GEC CCD. Of these spectra, 8 were taken at a
dispersion of 1.121\,\AA/pixel and the remaining 34 at 2.304\,\AA/pixel, including
several wide slit spectra. The corresponding wavelength ranges were 7662 --
8308\,\AA\ and 6951 -- 8280\,\AA. Table~\ref{syj:spec} gives a full journal of the
observations. A copper-argon/copper-neon arc spectrum was taken  at one- or
two-hourly intervals during the observations, to provide wavelength calibration
and to correct for instrumental  flexure and instability. Spectra of the
spectrophotometric standards HD 84937, Feige 34, BD+26$^{\circ}$2606 (Oke \& Gunn
1983) \nocite{okegunn83} and BD+25$^{\circ}$3941 were also  taken in order to
remove the telluric absorption features and correct for the instrumental
wavelength response. Eleven single red dwarfs in the spectral type range K7 to
M5.5 were observed at low resolution to allow spectral type calibration. The
details are given in Table~\ref{syj:red}.

\subsubsection{Photometry}

We also observed SY~Cnc photometrically on 4 nights, using a GEC CCD on the
Jacobus Kapteyn Telescope. On the first night only colours were observed, while a
long time series of I-band data was taken on each of the other 3 nights.
Table~\ref{syj:phot} gives a full journal of the observations. We achieved
significant overlap between the spectroscopic and photometric sets of observations
on 3 nights (nights 5, 6 and 7), allowing us to undertake flux calibration both by
using the simultaneous photometry and by using the wide slit spectra and the
standard stars. This enabled an interesting comparison of the two methods of
absolute flux calibration and also allowed us to check whether the slit losses
were ``grey'' or wavelength dependent.

\begin{figure*}
\includegraphics[width=14cm]{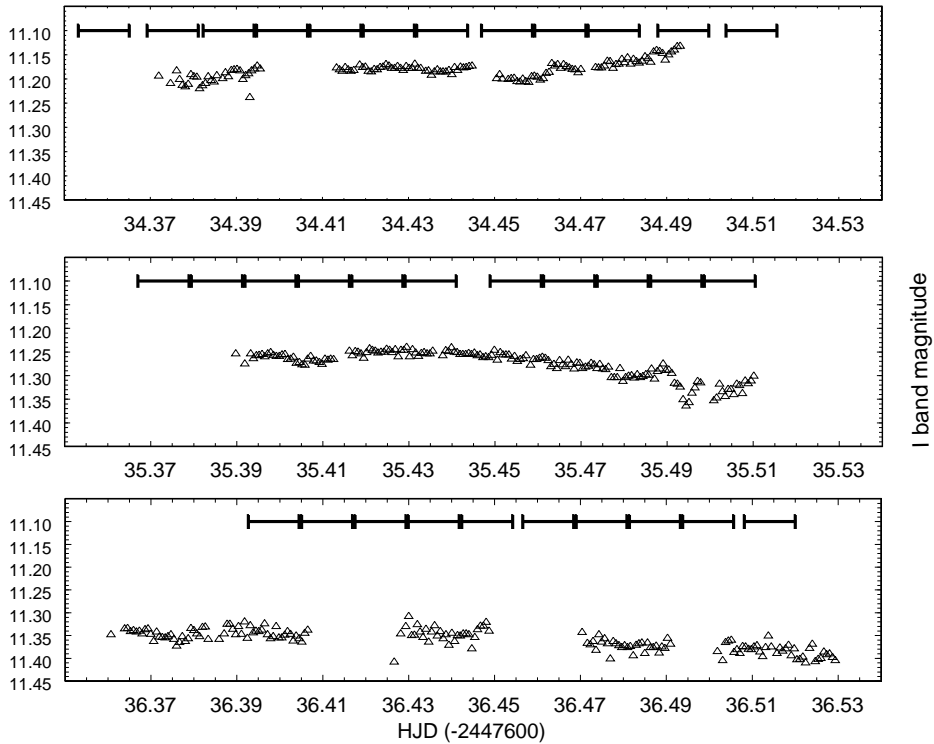}
\caption[The 3 nights of simultaneous photometric and spectroscopic data for
SY~Cnc] {The 3 nights of simultaneous photometric and spectroscopic data for
SY~Cnc (top: night 5; middle: night 6; bottom: night 7). The triangles are the I
band photometry points and the bars show the exposure durations of the spectra.}
\label{syphasecov}
\end{figure*}

\subsubsection{Ephemeris}

The ephemeris used for all the light and radial velocity curves was the one
given by Vande Putte et al (2003):

\[\mbox{HJD} = 244 7630.619 5678 + 0.380\,E \ . \]

The period is not very precisely known, but the precision cannot be improved by
our relatively short data run.

\subsection{Data reduction}

\subsubsection{Spectroscopy}

After the usual debiassing and flatfielding, the spectra were reduced using the
method of optimal extraction (Horne 1986). \nocite{horne86a} For the wavelength
calibration, sky emission lines were used to check the interpolation of the
calibration between arc exposures.  The continuum extinction was removed by using
the tabulated average values for the extinction per unit airmass on La Palma (D.
King 1988). \nocite{dking88} The atmospheric absorption was removed by using the
standard stars to create a template showing the fractional absorption in each
wavelength bin.  For each spectrum, the standard star spectrum taken nearest in
time to the object spectrum was used; Feige 34 was used for the wide slit spectra.
Finally, the spectra were flux calibrated using the standard star Feige 34.  This
provided relative fluxes, but not absolute flux calibration, because of slit
losses. The correction for slit losses is discussed in the next section.

\subsubsection{Photometry}

The CCD frames were first debiassed and flat-fielded and then the raw magnitudes
were obtained for SY Cnc and for the comparison and standard stars by using
aperture photometry. The wavelength-dependent extinction was taken from the tables
of Argyle \ea (1988) \nocite{argyle88} for La Palma; the correction for the I band
is 0.0174 mag. The wavelength independent dust component was removed by taking the
values of {\it total} V-band extinction measured every night at the Carlsberg
Automatic Meridian Circle and subtracting the tabulated wavelength-dependent
component for the V band. Photometric standard stars chosen from the list of
Argyle \ea (1988) were used to put the raw magnitudes on the absolute scale
defined by Cousins.

Fig.~\ref{syphasecov} shows the I-band photometry data for SY Cnc for the 3
nights, together with the exposure times of the spectra. The photometry data have
been corrected for all atmospheric effects, as described above, and show a very
clear decrease in brightness over the 3 nights. Table~\ref{syphot} gives a summary
of the magnitudes in all the observed wavebands during the run. Ritter \& Kolb
(2003) quote the V band magnitude to be about 10.9 in outburst, 12.2 in standstill
and 13.5--14.5 in minimum, so the system appears to have been on the decline from
an outburst at the time of our observations. None the less, as we shall see below,
we were able to detect the secondary star.

\begin{table}
\begin{center}
\begin{tabular}{|c||c|c|c|c|c|}
\hline
 Night  & \multicolumn{5}{c|}{ Cousins magnitude in band: } \\
 of run & U & B & V & R & I \\
\hline
 1 & 10.835 & 11.533 & 11.340 & 11.325 & 11.027 \\
 5 & 10.516 & 11.432 & 11.448 & 11.341 & $\sim$11.17 \\
 6 & 10.746 & 11.590 & 11.543 & 11.441 & $\sim$11.3 \\
 7 & 10.800 & 11.667 & 11.614 & 11.491 & $\sim$11.35 \\
\hline
\end{tabular}
\caption[]{Multi-colour photometry of SY~Cnc. Colours were measured at most once
per night, all long series of observations being made in the I band. Typical
exposure times were 32 seconds for the I and U bands and 16 seconds in the other
bands (where the instrument has a better response). Errors were typically
$\pm0.03$ magnitudes in all bands.} \label{syphot}
\end{center}
\end{table}

\section{Simultaneous Photometry or Spectrophotometry?}

Unlike the photometric data, the spectroscopic data are rather difficult to put on
an {\it absolute} flux scale. The main reason for this is the presence of {\it
slit losses}, caused by some of the light of the star failing to pass through the
slit. The amount of light which does pass through the slit depends on several
factors:

\begin{enumerate}
\item Slit Width --- We need to use a narrow slit to ensure that we get the
maximum spectral resolution possible with the instrumental setup being used, but
this means that some light is likely to miss the slit. Assuming other factors to
be constant, this will introduce a systematic error in the mean level of the
spectra.

\item Positioning --- If the object is not centred correctly on the slit,
then light will be lost. Any slight drift in the pointing etc. will move the
position giving rise to losses differing from frame to frame.

\item Seeing --- In poor observing conditions, the light from the object is
spread out more, and so a wider slit is  needed to gather all the available
photons. A further complication here is the possibility that the size of the
seeing  disc may vary with wavelength, so that rather than just producing
variations in mean level, the shape of the spectrum may be distorted.

\end{enumerate}

Fortunately, there are two methods of correcting for these losses, using
simultaneous observations either of the same object photometrically or of a nearby
non-variable comparison star spectroscopically.

\begin{description}

\item [{\bf Simultaneous Photometry.}] This method requires the use of a
second telescope to obtain the photometric data simultaneously with the
spectroscopic observations, but is slightly easier to perform than the comparison
star method. For every spectrum taken of the object, one or more simultaneous
photometry points in the waveband which matches the spectrum limits most closely
are needed. The average magnitude of these is converted to the relevant flux
density units and the average level of the spectrum is corrected to equal this
level. The spectra were all centred in the wavelength range 7600--7900\,\AA, so
the I-band data were appropriate for the calibration.

The first stage in the process was to find the average magnitude of all the
photometry points which overlap with each spectroscopic exposure. The Cousins I
magnitude was then converted to a flux density in mJy using the following relation
(based on data from Bessell~1979): \nocite{bessell79}

\[
\log_{10} f_{\lambda}  =  \left( \frac{I-16.016}{-2.5} \right) .
\]

The corresponding error relation is:

\[
\frac{\Delta f_{\lambda}}{f_{\lambda}}  = -0.921\Delta I .
\]

Our typical error of $\pm0.03$ magnitudes thus gave an error of approximately
$\pm3$ per cent in flux density. The major problem with this method is that although the
mean level of the spectra is corrected, it is not possible to correct for any
wavelength-dependent losses that would distort the spectral shape.

\vspace{0.1cm}

\item [{\bf Comparison Star Spectra.}] This method requires that a second
(non-variable) star is positioned on the slit together with the object. It also
requires that at least one spectrum is taken of the two stars and of a flux
standard, in photometric conditions, with a very wide slit to ensure that slit
losses can be accurately removed. Ideally, the wide slit object spectra should be
taken immediately before or after the wide slit standard star spectrum so that
atmospheric variations between the two exposures are kept to a minimum. This wide
slit standard spectrum is then used to flux calibrate the wide slit object and
comparison star spectra. The standard star used here is Feige 34, for which the
spectrophotometric data were obtained by Oke (1990)\nocite{oke90}.  The wide slit
spectra are assumed to show no slit losses and hence to provide an accurate
continuum level across the whole spectral range on an absolute scale. Note that
since a wide slit was used, the spectrum will not show line features very
accurately. Polynomials were now fitted to the continuum of both the wide and
narrow slit comparison star spectra. Each narrow slit object spectrum was
corrected for slit losses by multiplying it by the ratio of the comparison star
wide slit continuum fit to the appropriate comparison star narrow slit continuum
fit. Thereafter, the resulting spectra were multiplied by the ratio of a
polynomial fit to the tabulated spectrophotometric standard spectrum to a
polynomial fit to our observed wide slit standard spectrum. Both adjustments are,
of course, wavelength-dependent.

There is a problem with this method in that the need to place a comparison star on
the slit means that in general the observations are made with the slit not at the
parallactic angle (i.e. not vertical). This can be important since the atmosphere
(especially at large airmasses) acts as a weak prism and spreads the incoming
starlight slightly in the vertical direction. Having the slit vertical should
ensure that all the light spread in this way enters the slit. With the slit not
vertical, in order to obtain comparison star spectra, some of the light may miss
the slit. This problem is compounded if the wavelength response of the TV guider
differs from that of the detector as the telescope may appear to be perfectly
aligned but actually be missing most of the light at the desired wavelengths. In
addition, careful alignment of the slit is necessary to ensure that both stellar
images are central on it, otherwise slight drift in the telescope positioning
could lead to the recorded flux from the stars varying in relation to each other
as one may drift off the slit while the other moves further onto it. The use of
wide slit spectra should compensate for any errors due to the atmospheric prism
effect but not necessarily for bad alignment of the slit.

\end{description}

Either of these methods should absolutely calibrate the spectroscopic data, though
only the latter will correct for any wavelength dependence in the slit losses.
Usually, one or the other of these methods is used, but it was pointed out to the
authors (Dhillon~1990 and private communication) \nocite{dhillon90} that there
seems to have been no comparison between the two methods. As shown in
Fig.~\ref{syphasecov}, we have I band photometry simultaneous with 29 of the lower
dispersion spectra. All of these object spectra also had comparison star spectra
on the chip, and two of the spectra were taken with a wide slit, giving us all the
data necessary to compare the two methods of absolute calibration.

\begin{figure*}
\includegraphics[width=14cm]{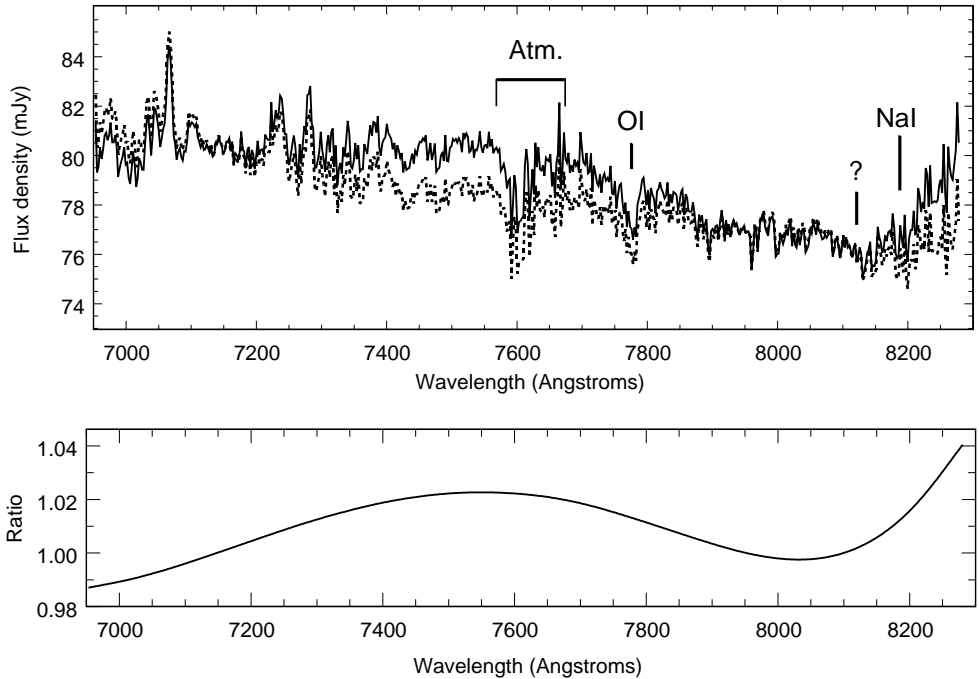}
\caption[Comparison of the two methods of absolute flux calibration.]{The top
panel shows the average SY~Cnc spectrum flux-calibrated using comparison star
spectra (dotted line) and simultaneous photometry (solid line). A few spectral
features and the atmospheric A band are marked, including the unidentified feature
(?) mentioned in section~\ref{SY:RVstudy}. The bottom panel shows the ratio of the
two spectra.} \label{sycnc_comp}
\end{figure*}

All 29 object spectra were flux calibrated using both methods described above, and
then averaged. Fig.~\ref{sycnc_comp} shows the two average spectra and a
comparison between them. The comparison is computed from the average of 29
spectra, each of which individually has an uncertainty of about 3 per cent, so we expect
a scatter of about $(3/\sqrt{29})$ per cent or $\sim0.5$ per cent in the average  and about
0.7 per cent (0.5$\times\sqrt{2}$) in the ratio. The comparison shows a slight
systematic difference of about 2 per cent between the two methods, but the ratio shows
only small ($\sim1$ per cent) variations from a straight line, which indicates that the
slit losses are almost wavelength independent. The curve appears to be very
smooth, but this is expected since the spectrophotometry process uses polynomials
fitted to the spectra to correct for slit losses. The origin of the 2 per cent
difference between the spectra is unknown, but is likely to stem from a small
error in the zero point calibration of the photometry. We also note that the
I band photometry is centred at a wavelength of around 9000\,\AA, whereas the
spectra are centred around 7600\,\AA; this might also lead to a small zero
point difference. The zero point error can be estimated from the equation
above to be about 0.02 magnitudes.

This result means that the two calibration techniques are essentially identical
(at least over this wavelength range) and so either may be used without
introducing unnecessary errors. Thus it is possible to choose whichever method is
most convenient in a given observing situation. The factors to be considered are:

\begin{enumerate}

\item Use of resources. The simultaneous photometry method normally requires
a second telescope. This will usually be difficult or impossible to achieve,
making the  comparison star spectrophotometry method the preferred choice in most
situations. What we have demonstrated here is that it is no less accurate.

\item Wavelength dependent slit losses. The above analysis shows the slit
losses to be `grey' over this wavelength range, but this may not necessarily be
true over other (particularly much wider) ranges and then spectrophotometry is
essential.

\item Differential refraction. Because the use of a comparison star does not
allow the slit to be used at the parallactic angle, there is always a residual
effect of differential refraction in spectrophotometric calibration, and it is
important to work as close to the zenith as possible.

\item Availability of a comparison star. The other big disadvantage of the
spectrophotometry method is that there must be a non-variable comparison star near
enough to the object to fit on the slit, and this may not always be true. In such
cases  simultaneous photometry must be used for absolute calibration. Our results
suggest that, at least for small wavelength ranges, this may be done without
distorting the resultant spectra.

\item For echelle spectra, the orders are generally too closely spaced on the CCD
to allow the spectrum of a comparison star to be included in the same exposure.
Simultaneous photometry is then essential, but in practice is hard to obtain if
one wishes to use 8-m class telescopes: for example, there is no readily available
photometric telescope available on the same site as the VLT.

\end{enumerate}

\section{Radial Velocity Study}
\label{SY:RVstudy}

\subsection{Sodium doublet velocities}\label{sodium}
\begin{table*}
\begin{center}
\begin{tabular}{|c|c||c|c||c|c|}
\hline
 HJD & RV (km\,s$^{-1}$) & HJD & RV (km\,s$^{-1}$) & HJD & RV (km\,s$^{-1}$) \\
\hline
 30.3762728 &     *** & 34.4254154 &  -30.08 & 35.4672223 & -103.21 \\
 30.3915678 &  -23.54 & 34.4378249 &   -8.94 & 35.4796619 &  -92.07 \\
 32.4379346 & -120.87 & 34.4528655 &    4.14 & 35.4921105 &  -90.94 \\
 32.4651007 & -162.32 & 34.4652682 &   55.63 & 35.5045500 &  -69.19 \\
 32.4779764 & -163.18 & 34.4776773 &   78.46 & 36.3985788 &  138.29 \\
 32.4913055 & -198.92 & 34.4938585 &  153.61 & 36.4110005 &  232.29 \\
 32.5039073 &  -98.88 & 34.5096814 &  153.37 & 36.4234270 &     *** \\
 32.5128255 &  -43.95 & 35.3729502 & -159.57 & 36.4358591 &     *** \\
 33.4014396 &  251.93 & 35.3853784 & -135.47 & 36.4482748 &  220.95 \\
 34.3591914 & -136.99 & 35.3978143 &     *** & 36.4624989 &  144.82 \\
 34.3752031 & -106.86 & 35.4102357 & -147.17 & 36.4749092 &     *** \\
 34.3881816 & -100.98 & 35.4226507 & -207.18 & 36.4873283 &   84.96 \\
 34.4005943 & -115.54 & 35.4350579 & -206.56 & 36.4997447 &  198.72 \\
 34.4130141 &  -92.53 & 35.4547941 & -108.45 & 36.5140711 &  181.12 \\
\hline
\end{tabular}
\end{center}
\caption[Radial velocity results for SY~Cnc with Gl~673 template.]{Radial velocity
results for SY~Cnc with the Gl~673 template, giving heliocentric julian date
(-2447600.0) at mid-exposure and measured heliocentric radial velocity for each of
the 42 spectra. The spectra marked with asterisks in the RV column were not used
in performing the fit.} \label{tab:sy_rvs}
\end{table*}

Given Shafter's~(1983) measurements of $86\pm9$\,km\,s$^{-1}$ for $K_1$ and
$1.13\pm0.35$ for the mass ratio, $q$, we would expect the secondary to be a late
G or early K type star, and to give a radial velocity curve having a $K_2$ of
$76\pm25$\,km\,s$^{-1}$. Our earliest red dwarf cross correlation template was
only K7, as the majority of the systems we were looking at have much shorter
periods than SY~Cnc and so the later templates were more generally useful. There
was thus a danger that we would be unable to obtain a good radial velocity curve
due to template mismatch, which has been indicated to be a significant source of
error by Tonry \& Davis~(1979). \nocite{tonrydavis79} The cross-correlation was
run using the full range of templates available (K7 to M5.5) to try to minimise
errors due to mismatch; it turned out that the best match was the K7
template, leaving some doubt at this stage about whether an earlier spectral 
type might provide a better fit (but see further discussion below, in the
context of our skew mapping).

All the initial cross correlation results were found to be very noisy, with a
blueshift of about 1200\,km\,s$^{-1}$ being indicated on several frames. All the
frames were examined by eye, and it was found that there was an absorption feature
centred at about 8150\,\AA, which could not be identified with any K or M dwarf
feature in the Turnshek {et al.} (1985) \nocite{turnshek85} spectral atlas. This
feature showed most strongly in those frames giving the very high blueshift, with
a strength roughly equalling that of the nearby 8190\AA\ sodium doublet (this
unidentified feature is marked on Fig.~\ref{sycnc_comp} with a question mark). In
the late K and early M dwarf spectra being used for templates, the 8190\AA \ \nai\
doublet is by far the strongest feature in the region used for the cross
correlations. The unidentified feature is of roughly the same width as the \nai\
doublet, and at roughly the right wavelength to give the 1200\,km\,s$^{-1}$
blueshift measured if it is picked up by the cross correlation routine in
preference to the doublet. This feature was assumed to be the source of the noise
and was masked out in later cross correlation attempts. Further discussion
of this feature is given in Section~\ref{UF}. 

\begin{figure}
\center
\includegraphics[width=7cm]{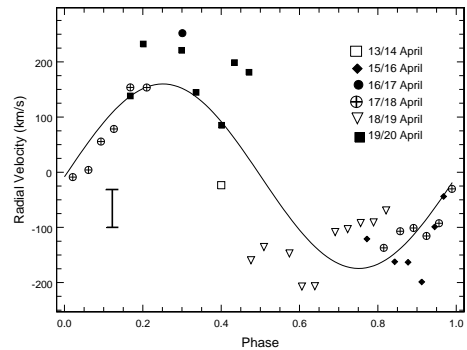}
\caption[Circular orbit fit to SY Cancri data.]{Circular orbit fit to SY~Cancri
data. The points are coded to show the night they were taken on and the bar in the
bottom left indicates the size of the RMS deviation from the fit.}
\label{sycnc_orbfit}
\end{figure}

After masking out the unidentified feature, the cross correlation procedure was
repeated, again using all the available templates, and was now found to give much
less noisy results. Any template spectrum of spectral type K7 to M1 gave fairly
good results in terms of noise level, though five very noisy spectra had to be
excluded from the process as they gave wildly varying shifts depending on the
template used. The template giving least noise was Gliese~673 (K7V), and
Table~\ref{tab:sy_rvs} lists the cross-correlation results from using this
template. The corresponding value of $K_{\rm 2}$ was found to be $168\pm15$\kms.
The tests given by Bassett (1978) \nocite{bassett78} were used to test whether
there was any significant ellipticity in the orbit. None was found, and
Fig.~\ref{sycnc_orbfit} shows the circular orbit fit corresponding to the
parameters in Table~\ref{sycnc_rv}. When the spectra are shifted and averaged
using these parameters the \nai\ doublet is significantly enhanced, and the value
of $K_{\rm 2}$ was used in a preliminary account of this work (Smith, Hawkins \&\
Mehes 1999), \nocite{smith99oxford} which reported a mass ratio of $0.51\pm0.07$.

\begin{figure}
\center
\vspace{5cm}
\includegraphics[width=7cm]{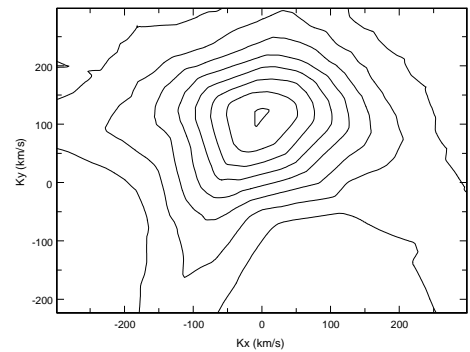}
\caption[SY Cnc skew map]{Skew map for SY Cnc, using Gliese 488 (M0$-$) 
as the template. Starting from the centre, the contours represent 99\%,
85\%, 71\%, 56\%, 42\%, 30\%, 15\%\ and 1\%\ of the peak height.}\label{skew}
\end{figure}

\begin{figure}
\center
\includegraphics[width=7cm]{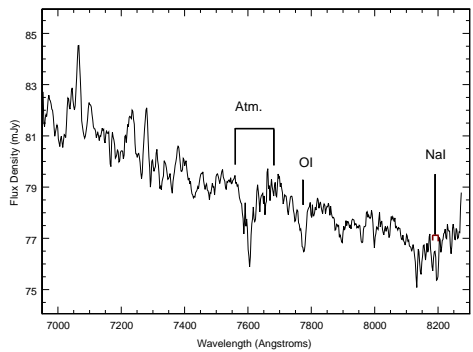}
\caption[Mean SY Cancri spectrum.]{Mean spectrum of SY~Cancri. The individual
spectra which make this up have been shifted to correct for the motion of the
secondary star before adding. A few spectral features and the atmospheric A band
are identified.} \label{sycnc_orbspec}
\end{figure}

\begin{figure}
\center
\includegraphics[width=7cm]{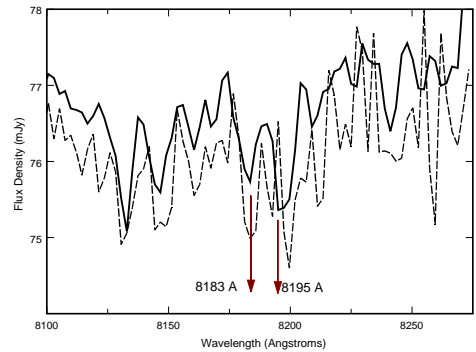}
\caption[detail]{An expanded view of the region of Fig.~\ref{sycnc_orbspec} 
near 8200\,\AA, to show the enhancement of the \nai\ doublet as a result of
correcting for the secondary motion. The dashed line shows the average without
correction, and the solid line shows the average spectrum after correction;
note that the three or four features in the unshifted average combine into
just two features, at the correct wavelengths for the \nai\ doublet.}
\label{detail}
\end{figure}

However, this radial velocity curve is still very noisy, as shown by the very 
large dispersion, and the large effect of
removing the unidentified feature makes the result questionable. We therefore
decided to adopt instead the technique of skew mapping (Smith \ea\ 1993a)
\nocite{skew93} which is designed to make optimum use of all the spectra.
Essentially, the technique consists of calculating all the cross-correlation
functions, as above, but then instead of using the individual velocity shifts the
next step is to use back-projection on the cross-correlation functions to produce
a map in velocity space. Each point in the map is the line integral over all the
cross-correlation functions following the sine curve with the particular amplitude
and phase at that point in the map. When calculating the cross-correlation
functions, we masked out all the spectral region shortward of 7800\,\AA, to 
avoid the \oxi\ absorption feature and the strong telluric A band around
7600\,\AA. We also had an optional mask for the region 8125 to 8175\,\AA, to
test the effect of the unidentified feature. The skew map was unaffected by
whether or not we masked out the region containing the unidentified feature,
showing the ability of skew mapping to screen out outliers.
 
The skew map (Fig.~\ref{skew}) has a very clear peak at the expected orbital 
phase, with no subsidiary noise peaks of comparable size. The skew map 
satisfied all checks recommended in Vande Putte et al.
(2003)\nocite{vandeputte03}.  These relate to the position of the peak of the skew
map and its intensity, to a recognisable sine pattern in the position of the
maxima of the cross-correlation functions, and to the resolved features in the
average shifted spectrum.  This produces the result in Table~\ref{sycnc_rv}.
Fig.~\ref{sycnc_orbspec} displays an orbital average spectrum corrected for this
secondary motion and again demonstrates an enhancement of the \nai\ feature (see 
Fig.~\ref{detail} for an expanded view of the region near the \nai\ doublet).
However, the deduced amplitude of the radial velocity curve is considerably less
than found from the straightforward cross-correlation study. The errors in that
study were re-examined, and it was found that they had been significantly
under-estimated, although the value of $K_{\rm 2}$ was confirmed. The error on
$K_{\rm 2}$ quoted in Table~\ref{sycnc_rv} is simply the formal statistical error
of the fit; a better estimate of the uncertainty in $K_{\rm 2}$ from the
simple cross-correlation technique is the dispersion
about the radial velocity curve. The dispersion is so large that the two results
for $K_{\rm 2}$ are in fact consistent, as can be seen in Table~\ref{sycnc_rv},
although the value from skew-mapping is believed to be considerably more reliable.

We note that a skew map with the M1.5 template Gl 205 also produces a significant
peak, albeit some 15 per cent lower (Vande Putte 2002)\nocite{vdp02}. With 
our other templates, the peak intensities of the maps are much lower, and in 
some cases the peaks lie at the wrong phase. We are therefore confident that 
the M0$-$ template provides the best fit amongst our observed templates.

Harrison et al. (2004)\nocite{harrison04} have suggested a spectral type of G2
for SY Cnc, based on the relative strengths of absorption lines in infrared 
spectra over a limited phase range (0.64 to 0.75). 
We therefore sought a G2 template in 
the literature, so that we could try a skew map for a wider range of templates.
We took two spectra, of spectral types G2V and K0V, from the STELIB
compilation of Leborgne et al. (2003)\nocite{leborgne03}, smoothed them
to the same resolution as our spectra of SY Cnc, and used them as templates. 
The resulting skew maps both had much lower intensity peaks than those for our
M0$-$ template, and the peaks were at the wrong phases. We are therefore very 
doubtful that the G2 spectrum found by Harrison et al. represents the
average spectrum of the secondary star. We speculate that it may arise from
a heated region on the leading hemisphere, which would be preferentially 
visible in the phase range of their observations. Unfortunately, our data
are not of high enough quality to test this by mapping the surface, but
other CVs certainly show such regions (e.g. Davey \&\ Smith 1992).

\begin{table}
\begin{center}
\begin{tabular}{c*{2}{r@{$\pm$}l}}
\hline
 & \multicolumn{2}{c}{\bf Gl 488} & \multicolumn{2}{c}{\bf Gl 673} \\
 & \multicolumn{2}{c}{M0-} & \multicolumn{2}{c}{K7V} \\ \hline
 RV shift fit: & \multicolumn{2}{c}{ } & \multicolumn{2}{c}{ } \\
 $\gamma$ &  \multicolumn{2}{c}{ } & -7.2 & 16 \\
 $K_{\rm 2}$ & \multicolumn{2}{c}{ } & 168.3 & 15 \\
 Dispersion ($\sigma$) & \multicolumn{2}{c}{ } & 68.9 & 8.9 \\ \hline
 Skew mapping: & \multicolumn{2}{c}{ }& \multicolumn{2}{c}{ } \\
 $\gamma$ & 32 & 13 &  \multicolumn{2}{c}{ } \\
 $K_{\rm 2}$ & 127 & 23 &  \multicolumn{2}{c}{ } \\ \hline
 $q$ & 0.68 & 0.14 & \multicolumn{2}{c}{ } \\ \hline
\end{tabular}
\end{center}
\caption[SY~Cnc radial velocity study results.]{SY~Cnc radial velocity study
results (km\,s$^{-1}$) for the template giving the skew map with the strongest
peak (Gl 488), and for the template giving the lowest dispersion radial velocity
fit (Gl 673). The results from the simple fit to the radial velocity curve
have such a large dispersion that they cannot be used. The last
line is the derived mass ratio using Shafter's $K_{\rm 1}$ value of
86$\pm$9\,km\,s$^{-1}$.} \label{sycnc_rv}
\end{table}

\subsection{The unidentified feature}
\label{UF}

The feature near 8150\,\AA\
could not be positively identified. We initially thought that it was an artifact
introduced by the extinction correction of the spectra, because the feature 
appears at roughly the same wavelength as the start of the strong atmospheric 
water feature which overlies the sodium doublet and which, on these fairly 
noisy spectra, might have been poorly removed. Certainly, the very strong 
atmospheric A band was not completely removed -- see Fig.~\ref{sycnc_orbspec}.
However, as indicated in the previous section, masking out the unidentified
feature has no effect on the result of skew mapping.

When skew mapping for the individual nights, it was noted that nights 6 and 7
yielded a $K_2$ and phase within 5\%\ of the original result for all three
nights together. Again, masking out the feature had no effect on this. Night
5 on its own, however, does not produce a credible map (the peak line integral 
is negative). Considering the three nights separately, it appears that the
strength of the unidentified feature decreases with time. On night 5, the 
feature was about twice the strength of the \nai\ doublet, on night 6 it was 
about the same strength as the doublet and on night 7 it had decreased to
about half the strength of the doublet. From this we surmise that the
feature is associated with the outburst, and the decrease in strength may be 
a sign of the disk gradually becoming optically thin.

\subsection{The masses of the components}
\label{masses}
The most important result in Table~\ref{sycnc_rv} is that the mass ratio is now
significantly less than unity and is thus completely consistent with the canonical
model for mass transfer in CVs. With this independent result for $K_2$, what can
we deduce about the masses of the two stars?

If we first make the common assumption that the secondary just fills its Roche
lobe, and lies on the main sequence, we can use Smith \& Dhillon's~(1998)
empirical \nocite{smith98} main sequence mass--period relation to derive the
following values of $M_{1}$ and $M_{2}$ by using our value of $q$:

\begin{eqnarray}
M_{1} & = & 1.54 \pm 0.40 \msun \nonumber \\ M_{2} & = & 1.04 \pm 0.11 \msun \ .
\nonumber
\end{eqnarray}


The most striking point about this is that the estimate for $M_{1}$ is above 
the maximum mass limit for a white dwarf, although the estimate for $M_{2}$ 
agrees well with that of Shafter (and is consistent with the spectral type of 
G2 claimed by Harrison et al. 2004). Furthermore, we note that the white dwarf 
mass is much higher than the average, $(0.84 \pm 0.29)\msun$, for Dwarf Novae 
above the period gap in Smith \& Dhillon's (1998) survey.  These facts suggest 
that at least one of the assumptions which led to Shafter's estimate of $M_2$ 
is incorrect.

However, given that we now have independent measures of both $K_{1}$ and $K_{2}$
we can make mass estimates without having to make the assumptions above. The only
assumptions we need to make are:


\begin{figure*}
\center
\includegraphics[width=14cm]{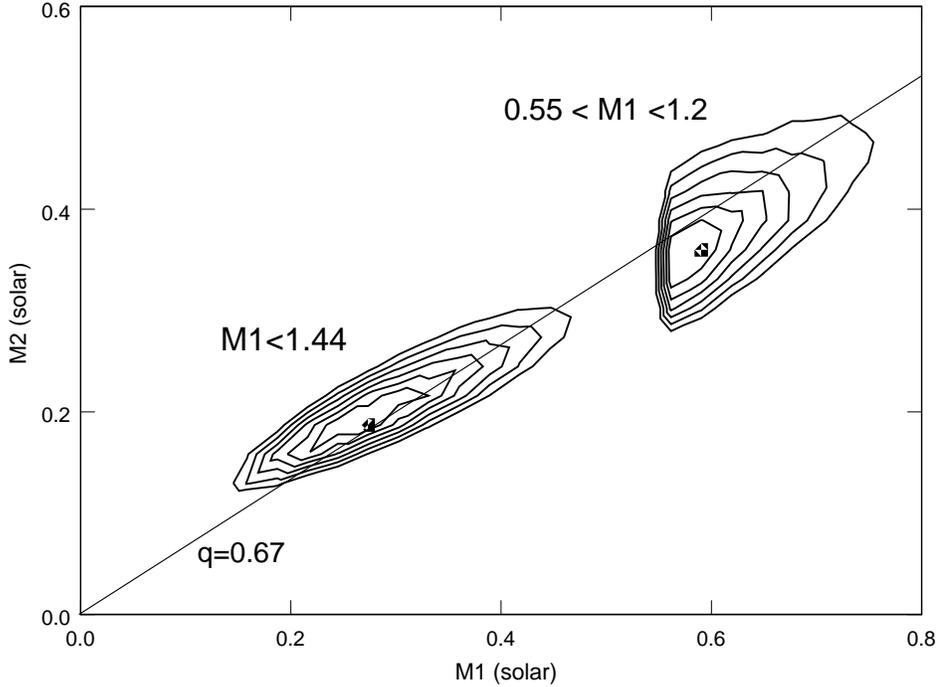}
\caption[Mass solutions for SY~Cancri.]{Mass solutions for SY~Cancri. 
The two contour plots represent the frequencies of successful outcomes from 
a $2\times10^7$ trial Monte Carlo sampling scheme where $i$ is uniformly 
distributed over $i=0-71^\circ$, and the primary mass is subject to the two
constraints $M_1 \le 1.44\msun$ and $0.55 \le M_1/\msun \le 1.22$. The diagonal
line labelled $q=0.67$ represents the stability line for secondary masses less
than 0.4342\,M$_\odot$. The lowest contour is at half the maximum frequency.
The corresponding plots for a uniform distribution of $\cos i$ are very
similar.} 
\label{sycnc_massfn}
\end{figure*}

\begin{itemize}

\item The stars are in circular keplerian orbits.

\item The maximum possible mass for a white dwarf is $\sim$1.44\,M$_{\odot}$

\item We would expect to see some sort of primary eclipse if the inclination
were higher than about $71^{\circ}$ with this mass ratio (Bailey, 1990).
\nocite{bailey90}

\end{itemize}

Horne, Wade and Szkody (1986) \nocite{horne86b} describe a Monte Carlo method for
inferring the component masses in such an instance.  This relies on having
reliable values for both $K_{1}$ and $K_{2}$, and using the mass functions:

\[
{\mathop {M_{1}} \nolimits_{}} \sin ^{3}i = {\frac{{PK_{2} (K_{1} + K_{2}
)^{2}}}{{2\pi G}}}
\]

\[
{\mathop {M_{2}} \nolimits_{}} \sin ^{3}i = {\frac{{PK_{1} (K_{1} + K_{2}
)^{2}}}{{2\pi G}}} \ .
\]

They sample the inclination at random from a uniform distribution over the
interval [$0,i_{\rm max}$]\footnote{They found that sampling from a uniform
distribution for $\cos i$ rather than $i$ had little effect.}. The values of
$K_{1}$ and $K_{2}$ are sampled at random from a normal distribution whose average
is the value of the velocity, and whose standard deviation is the reported error.  Using
the relations above, $M_{1}$ and $M_{2}$ are calculated.  If $M_{1}$ exceeds 1.44
solar masses, the outcome is rejected. The ensuing distribution for accepted
outcomes is used to calculate the component masses.  A similar calculation was
undertaken here, with random sampling from a uniform $i$ over the
interval [$0,i_{\rm max}$]. The results appear in Fig.~\ref{sycnc_massfn}.

From this, we obtain the following values for the system parameters, together 
with the very similar results (in brackets) from using a uniform distribution 
of $\cos i$:
\begin{eqnarray*}
   M_{1} & = & 0.27^{+0.20}_{-0.13}\, \hbox{M}_{\odot} 
\hspace{1cm} (0.25^{+0.21}_{-0.10}\, \hbox{M}_{\odot})   \\
\vspace{1mm} M_{2} & = & 0.19^{+0.12}_{-0.06}\, \hbox{M}_{\odot} 
\hspace{1cm} (0.19^{+0.11}_{-0.06}\, \hbox{M}_{\odot})   \\
\vspace{1mm}
   q     & = & 0.68^{+0.12}_{-0.23} \hspace{1.6cm} (0.76^{+0.02}_{-0.35})   \\
\vspace{1mm}
   i (^{\circ})  & = & 68^{+3}_{-38} \hspace{2cm} (63^{+8}_{-32}) .  
\end{eqnarray*}


If the primary is assumed to be a CO white dwarf with a mass range of
$0.55M_{\odot}$ to $1.22M_{\odot}$, then these values become:
\begin{eqnarray*}
   M_{1} & = & 0.59^{+0.16}_{-0.04}\, \hbox{M}_{\odot}
\hspace{1cm} (0.56^{+0.15}_{-0.02}\, \hbox{M}_{\odot})  \\
\vspace{1mm}   M_{2} & = & 0.36^{+0.14}_{-0.08}\, \hbox{M}_{\odot} 
\hspace{1cm} (0.33^{+0.13}_{-0.06}\, \hbox{M}_{\odot})  \\
\vspace{1mm}   
q     & = & 0.61^{+0.02}_{-0.21} \hspace{1.6cm} (0.59^{+0.10}_{-0.20})  \\
\vspace{1mm}   i (^{\circ})    & = & 52 \pm 19 \hspace{1.8cm} (55^{+16}_{-20}). 
\end{eqnarray*}

These results are also plotted in Fig.~\ref{sycnc_massfn}. Each value 
corresponds to the peak of the relevant distribution. Note that
the value of $i$ is significantly larger than estimated by Shafter (1983).

The two sets of parameters are distinct, so we need some way of choosing
between them. It turns out that they are in a mass ratio range that is close 
to the condition for dynamical instability for mass transfer. According to 
Politano (1996), for $M_2 \le 0.4342$\msun\ mass transfer occurs on a 
dynamical timescale for $q \ge 0.67$. Our first set of parameters therefore 
correspond formally to unstable mass transfer (especially for the uniform
$\cos i$ case), although the errors are large enough to make a firm conclusion 
impossible. However, the second set of parameters appears to be more 
definitely below the stability line, except possibly for the uniform $\cos i$
case. We therefore prefer the second set of parameters and will adopt
them in the later discussion.

\section{Flux variations around the orbit}
\subsection{I-band photometry}
\label{ibandphot}

\begin{figure*}
\includegraphics[width=14cm]{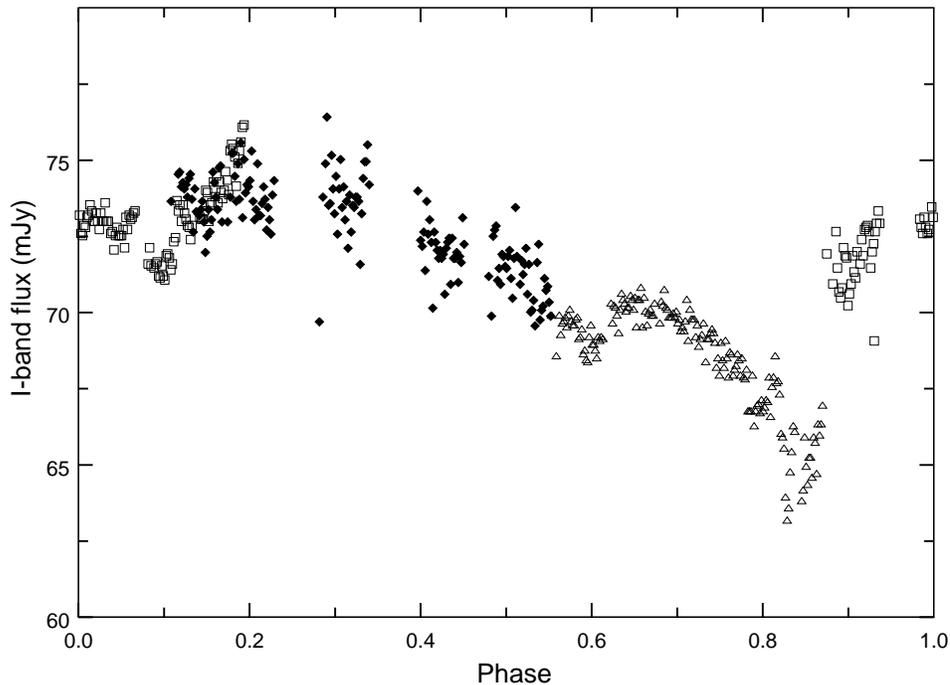}
\caption[]{The variation around the orbit of the I-band flux, adjusted to the flux
scale of night 7. Different nights are indicated by different filled symbols:
 square -- night 5; triangle -- night 6; diamond -- night 7. Phase zero corresponds
to the inferior conjunction of the red dwarf. } \label{orbfluxcurve}
\end{figure*}

Because of the relatively long orbital period, none of the three nights of
photometric data covers a complete range of phase and it is necessary to combine
the three nights in order to discover how the I-band flux varies around the
orbit. If the system were in quiescence, this would be relatively
straightforward. Because it is in decline from outburst, it is necessary to
devise some way of taking out the night-to-night variations before the orbital
variations are sought.

Inspection of Fig.~\ref{syphasecov} and Table~\ref{syphot} shows that the decrease
in brightness is less than 0.1 mag from one night to the next. The typical
duration of the observing run on a single night was less than 4 hours, so the
general decrease in brightness during one night's observation was less than 0.017
mag, smaller than the typical uncertainty of 0.03 mag in an individual point. We
therefore initially took out the night-to-night variations simply by computing the
average magnitude for each night and subtracting the appropriate night's average
magnitude from each observed point. We then plotted the magnitude differences as a
function of orbital phase, and it became apparent that the three nights had very
little overlap in phase and that there was considerable variation around the
orbit. This meant that even in quiescence we would not expect the average
magnitude to be the same for each night, as we had tacitly assumed.

Although the three nights cover almost disjoint ranges of phase, there is a very
slight overlap, especially between nights 5 and 7. We therefore adopted a
different strategy. We first converted the magnitudes to fluxes, and then adjusted
the mean flux for the data on nights 5 and 6 in such a way that the fluxes in the
overlapping ranges of phase matched. Because there is considerable scatter in the
flux curve, and the phase overlap is very small, this procedure is not very
precise. However, it produces a plausible looking light curve, which is presented
in Fig.~\ref{orbfluxcurve}. The total range of the variation is about 0.2 mag, or
about 20 per cent in flux. Because the simultaneous spectroscopic data cover a slightly
larger phase range, they were used to confirm that the flux adjustment had been
done consistently. We fitted a 3rd-order polynomial to the continuum of each
spectrum, between 7875\,\AA\ and 8075\,\AA\ to avoid the strongest lines, and used
these polynomial fits to produce an indicative continuum flux for each spectrum.
We then adjusted the mean flux for these data on nights 5 and 6 so that the fluxes
in the somewhat larger overlapping phase ranges matched; the resulting continuum
light curve is broadly similar to the I-band light curve, and in particular
confirms the fitting between the different nights. However, there is no large
line-free region of continuum to which to do the fitting, so the continuum light
curve is not a very reliable indicator of the more detailed flux variations around
the orbit and we do not present it here.

The main feature of the I-band curve is the drop to a relatively deep minimum at
about phase 0.85, and an apparently rapid recovery from it. This looks almost like
an eclipse, but the phase is wrong, the curve is rather asymmetric and the
inclination of the system is too low. The light curve also displays a surprisingly
large scatter about the mean curve, especially on nights 7 (diamonds) and 6
(triangles). Because we only just cover the entire phase range, it is not clear
whether either of these features would repeat from one cycle to the next. The
somewhat different behaviour in the phase range 0.1 to 0.2 on nights 5 and 7
suggests that they do not, and that we may simply be seeing fluctuations in the
disc brightness as it adjusts its structure after the outburst. Even the minimum
at phase 0.85 may arise from such fluctuations, and we refrain from speculating on
its significance until a better light curve is available.




\subsection{Spectroscopic variations: the \oxi\ line}
\label{oiflux}

The \oxi\ triplet absorption feature at 7773\AA\ is generally believed to arise in
the accretion flow, and is usually strong in nova-like systems (e.g. V1315 Aql;
Smith \ea\ 1993b).\nocite{smith93} However, it is also reasonably prominent in
this dwarf nova (Fig.~\ref{sycnc_comp}), presumably because the disc is still in a
bright post-outburst state, and we may use it to probe the nature of the accretion
flow.

\begin{figure}
\center
\includegraphics[width=7cm]{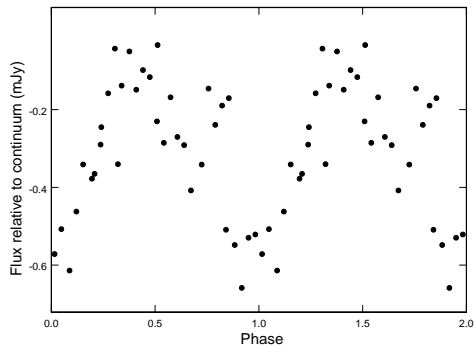}
\caption[]{The integrated flux (mJy) in the \oxi\ absorption triplet as a function
of orbital phase.} \label{oxiflux}
\end{figure}

We therefore measured the equivalent width of the absorption feature as a
function of orbital phase. Because some of the individual spectra are rather
noisy, the measurements were made several times and the results averaged. In
addition, we also measured the flux deficit for the feature by fitting the
entire continuum by a low order cubic spline function, using that to normalise
the spectrum, subtracting 1 from the resulting spectrum and integrating over
the (negative) absorption feature. All the results were consistent. The flux
deficit curve is shown in Fig.~\ref{oxiflux}, plotting two orbital cycles for
clarity.

The curve appears to be roughly sinusoidal, with a maximum absorption strength
around phase 0.9--1.0 and a broad minimum around phase 0.4. There also appears to
be a poorly defined but strong dip in absorption line strength around phase 0.8;
it is present in both the flux and equivalent width plots, so it is probably real.
However, it is not what is expected from a symmetrical disc. Since it is only
defined by four points, we do not try to interpret it.


If the absorption feature arose in the bright spot, it would be expected
to be at maximum strength when the observer was looking most directly at the
bright spot, which would be around phase 0.9. Because the absorption feature
is strong, we tried to verify this interpretation by measuring its radial
velocity curve. The template used for cross-correlation was formed by
taking one of the spectra in which the \oxi\ line was strongest and replacing
the spectrum outside the line by a smooth fit to the continuum, which was then
normalised to unity. However, the radial velocity curve obtained from standard
cross-correlation analysis was too noisy for any conclusions to be drawn.

We then attempted a skew-mapping approach, as described above for the \nai\
feature. This produced the interesting result that the \oxi\ feature appears to be
formed very close to the centre of mass of the system: the skew map is consistent
with the peak being exactly at the centre of mass, but the peak is broad enough to
be consistent also with a position rather closer to the bright spot; formally,
$K_{\rm x} = +1 \pm 42$\kms, $K_{\rm y} = -8 \pm 19$\kms. When interpreted as a
radial velocity curve, the amplitude is $8\pm26$\kms and
the zero-crossing from positive to negative velocities is at phase
$0.98^{+0.24}_{-0.20}$, which is consistent with its tracking the motion of the
bright spot.

We therefore conclude that the \oxi\ feature comes primarily from the
bright spot.

\section{Conclusions}

Our measurement of the secondary star radial velocity curve for SY~Cancri leads us
to conclude that the mass ratio in this system is indeed {\it less} than unity, as
is the case for most CVs. This result disagrees with the work of Shafter~(1983),
and must mean that one or more of his assumptions is incorrect for SY Cnc. Despite
the fact that it is now clear that {\em on average} the secondary stars of CVs
cannot be distinguished from main-sequence stars (Warner 1995, Sections 2.8 \&\
2.9; Smith \&\ Dhillon 1998), the most likely assumption to be wrong {\em for an
individual star} is the use of a main sequence mass-radius relation. Our results
show that, with the long period of the system and the observed mass ratio, the
secondary star cannot be on the main sequence unless the primary is a neutron star
rather than a white dwarf (or at least is an unusually massive white dwarf). 
If it were a neutron star, we would expect the
system's behaviour to be quite different from that observed. A very massive 
white dwarf might just be accommodated. However, a main
sequence Roche-lobe filling star would have a mass of 1.04\msun, which would
correspond to a spectral type of about G2, very different from the M0$-$ 
suggested by the skew mapping (see Section~\ref{sodium}).

Our conclusion therefore is that the secondary star in SY~Cancri is 
substantially lighter than a main-sequence star of the same size (or 
alternatively much larger than a main-sequence star of the same mass). From 
values in Allen~(1973, p.209) we find that a \nocite{allen73} main-sequence 
M0 star should have a mass of approximately $0.47$\,M$_{\odot }$; the same
mass was quoted by Martin (1988, Table B1)\nocite{martin88} for a spectral
type M0.5$\pm1$.  This is just about consistent within the errors with our 
preferred value of $0.36^{+0.14}_{-0.08}$\msun, which would correspond to a 
spectral type of about M2.5 according to Martin's Table B1. However, a 
main-sequence star of that mass would have a radius considerably less than 
the Roche lobe radius. The star does fill its Roche lobe and must therefore 
be much larger than its main-sequence radius. Normally, this would imply 
a star that was also much cooler than the spectral type corresponding to 
its mass, but this star is if anything slightly hotter than expected.

We therefore seem to have a star that is both considerably larger than a 
main-sequence star of its mass, and also slightly hotter (at M0) than 
the spectral type (M2.5) expected from its mass. It must therefore be 
considerably more luminous than a main-sequence star of 0.36\msun.
At a period of over 9 hours, the secondary may well be slightly evolved, 
which could account both for the larger radius and the larger luminosity, 
although the low present-day mass suggests that the evolution took place long
ago and that the system is very old. The radius is also expected to be 
somewhat enlarged compared to the main-sequence value by the star being out 
of thermal equilibrium. Irradiation of the secondary
by the primary/disc may also cause some expansion.

Irradiation can show up as a distortion of the radial velocity curve
(e.g. Davey \&\ Smith 1992). \nocite{davey92} There seems to be no sign of 
that, since the RV curve is not significantly elliptical. However, the RV 
curve is also very noisy, and Catal\'an, Schwope \&\ Smith (1999) 
\nocite{catalan99} show that if the irradiation is symmetrical about the 
line of centres there may be significant heating without there being a 
significant eccentricity in the orbital solution. If irradiation is important, 
the value of $K_2$ will need some correction. However, with a relatively long 
period, and correspondingly relatively large separation, the irradiation is 
very unlikely to affect the value of $K_2$ by enough to change our main 
conclusions that the secondary star in SY Cnc is substantially less
massive than found by Shafter (1983) and departs markedly from a 
main-sequence mass-radius relationship.


 \subsection*{Acknowledgments}

Nigel Hawkins was supported by a SERC studentship. The data reduction and analysis
were carried out at the Sussex node of the PPARC Starlink Project. The Jacobus
Kapteyn and Isaac Newton Telescopes were operated by the Royal Greenwich
Observatory at the Observatorio del Roque de los Muchachos of the Instituto de
Astrof\'{\i}sica de Canarias. We are grateful to Drs Vik Dhillon and Derek Jones
for assistance with the observing.



\bibliography{jrnls,sycncpop}

\bibliographystyle{mn}


\bsp

\end{document}